\documentclass{article}

\PassOptionsToPackage{square,numbers}{natbib}
     \usepackage[final]{neurips_2020}

\usepackage[utf8]{inputenc} 
\usepackage[T1]{fontenc}    
\usepackage{hyperref}       
\usepackage{url}            
\usepackage{booktabs}       
\usepackage{amsfonts}       
\usepackage{nicefrac}       
\usepackage{microtype}      
\usepackage{graphicx} 
\usepackage{amssymb} 
\usepackage{footmisc} 
 
\title{Assessing the Quality of Gridded Population Data for Quantifying the Population Living in Deprived Communities}

\author{%
Agatha C. Hennigen de Mattos \\
School of Computer Science\\
University College Dublin\\
Ireland \\
\texttt{agatha.hennigendemattos@ucdconnect.ie} \\
\And
Gavin McArdle\\
School of Computer Science \\
University College Dublin\\
Ireland \\
\texttt{gavin.mcardle@ucd.ie} \\
\AND
Michela Bertolotto \\
School of Computer Science \\
University College Dublin\\
Ireland \\
\texttt{michela.bertolotto@ucd.ie} \\
}

\begin{document}

\maketitle

\begin{abstract}
Over a billion people live in slums in settlements that are often located in ecologically sensitive areas and hence highly vulnerable. This is a problem in many parts of the world, but it is more prominent in low-income countries, where in 2014 on average 65\% of the urban population lived in slums. As a result, building resilient communities requires quantifying the population living in these deprived areas and improving their living conditions. However, most of the data about slums comes from census data, which is only available at aggregate levels and often excludes these settlements. Consequently, researchers have looked at alternative approaches. These approaches, however, commonly rely on expensive high-resolution satellite imagery and field-surveys, which hinders their large-scale applicability. In this paper, we investigate a cost-effective methodology to estimate the slum population by assessing the quality of gridded population data. We evaluate the accuracy of the WorldPOP and LandScan population layers against ground-truth data composed of 1,703 georeferenced polygons that were mapped as deprived areas and which had their population surveyed during the 2010 Brazilian census. While the LandScan data did not produce satisfactory results for most polygons, the WorldPOP estimates were less than 20\% off for 67\% of the polygons and the overall error for the totality of the studied area was only -5.9\%. This small error margin demonstrates that population layers with a resolution of at least a 100m, such as WorldPOP’s, can be useful tools to estimate the population living in slums\footnote{Code available at: \href{https://github.com/ml-labs-crt/popgrids}{https://github.com/ml-labs-crt/popgrids}\label{footnote}}. \end{abstract}
\section{Introduction}
More than half of the world population live in cities and it is estimated that over a billion live in slums \cite{habitatTrackingProgressInclusive2018a}. Though many definitions of slums exist \cite{lilfordBecauseSpaceMatters2019}, in this paper the term “slum” refers to material conditions of poverty, such as lack of access to public services, water and sanitation for example, and lack of house durability. This is a problem in many parts of the world, but it is more prominent in low-income countries, where in 2014 on average 65\% of the urban population lived in slums, as opposed to 44\% in lower-middle-income and 24\% in upper-middle-income countries \cite{worldbankPopulationLivingSlums}. Slums are often located in areas that are ecologically sensitive, critical for biodiversity, close to water bodies or important for flood protection and, as a consequence, a city’s risks and vulnerability are highly concentrated in these settlements \cite{satterthwaiteBuildingResilienceClimate2020a}. In the aftermath of a disaster, uncertainty over population figures and demographic information constitutes one of the main barriers to accurate needs assessment \cite{nojiEstimatingPopulationSize2005} and the absence of such data makes it impossible to adequately measure the impact of the disaster. Hence, to build resilient communities, it is essential that slums and their population are mapped and that their living conditions are improved \cite{satterthwaiteBuildingResilienceClimate2020a}. In this context, the United Nations ratified goal 11.1 of the Sustainable Development Goals, which states that adequate, safe and affordable housing should be accessible for all and that slums should be upgraded by 2030 \cite{habitatTrackingProgressInclusive2018a}. Countries, therefore, are expected to be able to report the progress towards this goal, which includes providing reliable disaggregated data about the percentage of the population living in slums at national and sub-national levels \cite{habitatTrackingProgressInclusive2018a}. However, most of the information about the population living in these areas comes from census data, which is only made available at aggregated levels \cite{klemmerPopulationMappingInformal2020}, and updated at lengthy intervals, making the information obsolete and unsuitable to capture the dynamism of slums \cite{mahabirCriticalReviewHigh2018a}. Moreover, some governments are reluctant to recognize slums, which are often excluded from official statistics \cite{habitatTrackingProgressInclusive2018a}. As a result, other approaches to provide the percentage of the population living in slums have been developed, such as using high-resolution satellite imagery and ground-truth field surveys to train convolutional neural networks to classify slum imagery patches according to their residential occupancy levels \cite{klemmerPopulationMappingInformal2020}. Yet, the cost for acquiring and processing high-resolution imagery as well as conducting field surveys hinders the application of these methods at scale by governments in low-income countries or not-for-profit organizations. An alternative would be to map areas using freely available satellite imagery and employ other free sources of population data to quantify the number of slum dwellers. In this work, we investigate this latter cost-effective approach by assessing the suitability of gridded population datasets to estimate the population living in areas that have been mapped as slums in the metropolitan region of São Paulo, Brazil. More specifically, we compare the population estimates provided by WorldPOP and LandScan's gridded population data with population data obtained through field surveys. In the next section, the methodology is described. Section 3 discusses the results and Section 4 presents conclusions and future work.\par
\section{Methodology}
Gridded population data are freely available and are usually produced by private companies or research institutions using machine learning, statistics and census data. In this study, we evaluate two of these population layers, WorldPOP and LandScan, as they are the only ones available for 2010, the year that our ground-truth data was collected. The WorldPOP data were developed with the support of many organizations \cite{sorichettaHighresolutionGriddedPopulation2015}, while the LandScan data were created by the Oak Ridge National Laboratory, sponsored by the U.S. Department of Energy. The gridded population data consist of a raster file with values that represent the number of people living in a small square area, also called a grid, with a resolution that can go from a 3 arc-second (100 m), in the case of WorldPOP, to 30 arc-second (1 km) for LandScan (see Figure \ref{Grids} for an example of the data for our area of interest). \par

\begin{figure}[ht!]
\centering
\includegraphics[width=0.55\columnwidth]{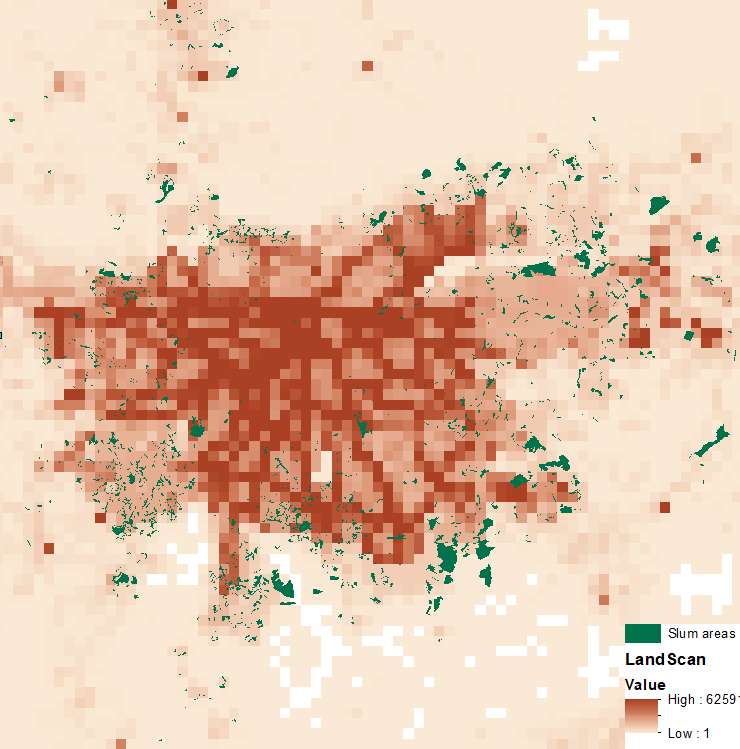} 
\caption{LandScan Gridded Population Data (2010) for the metropolitan region of São Paulo and georeferenced polygons of deprived areas (ground-truth data).}
\label{Grids}
\end{figure}

Gridded population data are generated by complex models that use a top-down approach to disaggregate census data. LandScan uses official census data and projections calculated by the US Census Bureau International Programs and WorldPOP utilises the Gridded Population of the World 4.10 \cite{sedacGriddedPopulationWorld}. Additionally, WorldPOP offers two options of gridded population layers, one that is adjusted using United Nations statistics and one that is not adjusted. In this work, we chose the latter version as our ground-truth is from the same year that the census was conducted. As mentioned previously, census data and projection statistics are only an input to these models that also employ many other parameters such as information on roads, land cover, built structures, cities or urban areas, infrastructure, environmental data, protected areas, water bodies and, in the case of WorldPOP, night-time lights \cite{trendsLeavingNoOne2020} to disaggregate census population counts into grids. WorldPOP uses random forests to select the geospatial features that will be used in the model and also makes its code open source \cite{sorichettaHighresolutionGriddedPopulation2015}. On the other hand, LandScan’s model is not completely open \cite{trendsLeavingNoOne2020} and is adapted to each region to account for specific geographical nature and data conditions \cite{landscanFrequentlyAskedQuestions}. \par
\begin{figure}[ht!]
\centering
\includegraphics[width=0.95\columnwidth]{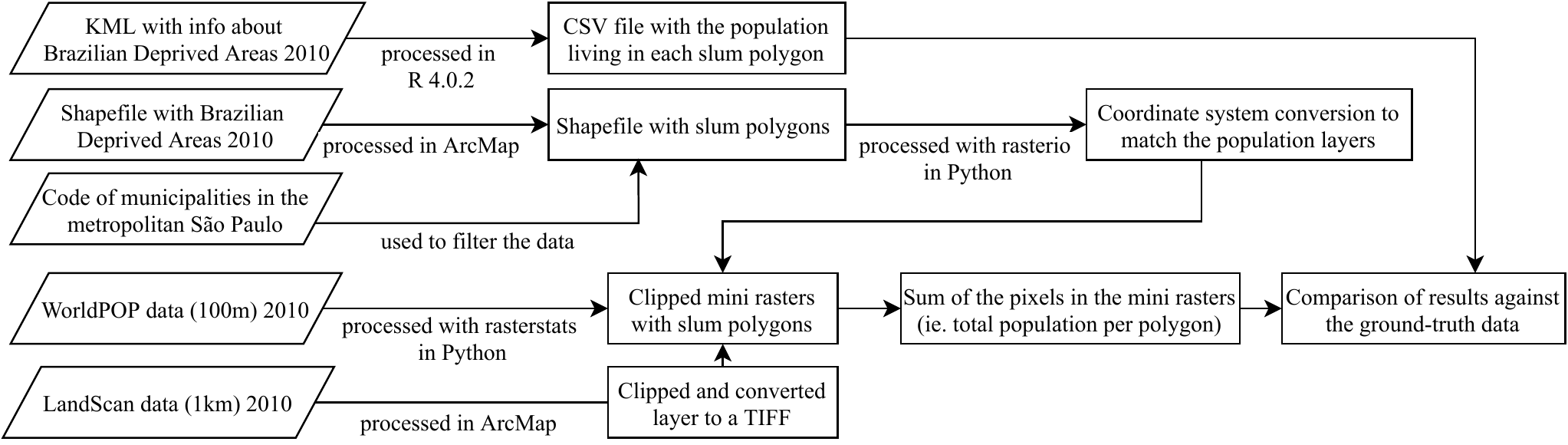} 
\caption{Methodology used to extract the population of each slum area in the metropolitan region of São Paulo.}
\label{Diagram}
\end{figure}

The ground-truth data used to assess the accuracy of the two gridded population data can be seen in Figure \ref{Grids} and is composed of 1,703 polygons that were mapped as deprived areas in the 2010 Brazilian census through field-surveys \cite{ibgeAglomeradosSubnormais2010}. Each polygon has information about the name of the settlement, geographical coordinates of the area and the number of inhabitants. The polygons are contained in the metropolitan region of São Paulo, which in 2010 was home to 18.9\% of the country’s population living in slum-like conditions. The criteria used for mapping these communities was that the majority of households in the surveyed area lacked access to at least one basic public service, tenure of the land, either at the moment of the survey or during the last 10 years, and that the area did not meet construction standards. As mentioned in the introduction, many definitions of slums exist and, as a consequence, any criteria can introduce biases by including or excluding households depending on their housing conditions. For instance, the criteria used in our ground-truth data does not include areas in which only a few households lack access to public services.
Figure \ref{Diagram} summarizes the methodology used in our experiments and our code is available online\footref{footnote}. Firstly, the ground-truth data was pre-processed to extract  population information, filter our region of interest and convert the coordinates to the same coordinate system of the gridded population data. Secondly, population layers were clipped in mini-rasters matching the location of the polygons in the ground-truth data. Lastly, a comparison of the results against the ground-truth data was obtained.

\section{Results and Discussion}
The comparison of the estimates obtained with WorldPOP \cite{sorichettaHighresolutionGriddedPopulation2015} and LandScan data \cite{brightLandScan20102011} and the ground-truth data shows that satisfactory results could not be obtained using LandScan’s population layer, but that WorldPOP produced an estimate of the population living in slums that was only 5.9\% less than the official data (it estimated 2,035,866 people living in the deprived areas whereas the ground-truth data indicate 2,162,368). When it comes to the performance at polygon level, the WorldPOP estimated the population of 67\% of the polygons within a relative error between 0\% and $\pm$20\%, and 24\% of the polygons within an error ranging from $\pm$20 to $\pm$100\% (see Table \ref{Grids}). The relative error was calculated for each polygon by dividing the difference between the estimated population and the ground-truth population by the ground-truth population. LandScan’s poor performance can be attributed to its coarse resolution in comparison to the area of the polygons, which leads to a situation in which any strategy to count the values of the pixels can produce errors. That is, if pixels that only partially overlap the polygons are included, they produce an overestimation of the population whereas if they are not included, many polygons’ population estimates result in null values. \par
Even though the literature of the use of gridded population data to estimate slum dwellers is, to the best of our knowledge, non-existent, WorldPOP results are in line with the work of \citet{sliuzasAssessingQualityGlobal2017}. \citet{sliuzasAssessingQualityGlobal2017} found that Global Human Settlement Layer products tended to underestimate the built-up area in deprived areas such as slums in Kampala, Uganda, when compared to the built-up area obtained using high-resolution satellite images. Since WorldPOP also employs built-up structures in their modelling, it is possible that the reason behind the underestimation is associated with missing built structures in slum areas. Indeed, WorldPOP has already been found to overlook built structures in rural areas \cite{trendsLeavingNoOne2020}. Other gridded population data, such as the High Resolution Settlement Layer \cite{facebookHighResolutionSettlement2016}, uses satellite imagery with higher resolution than WorldPOP in their model, which could potentially lead to better results. However, such datasets are not available for the year 2010, when the ground-truth data was collected, and for this reason were not included in this analysis. Regarding the differences in relative error of certain polygons, for instance 22 polygons had errors superior to 100\%, further examination using high-resolution satellite imagery can provide insight into the reasons behind it. On the positive side, the methodology proposed in this paper can easily be replicated for other regions. Still, both WorldPOP and LandScan make use of a top-down approach that does not interact with the local communities, something that has been argued by researchers to be desirable \cite{klemmerPopulationMappingInformal2020}. 
\begin{table}[]
\centering
\small
\begin{tabular}{ccccc}
& \multicolumn{2}{c}{WorldPOP} & \multicolumn{2}{c}{LandScan} \\ \hline
Relative Error & \# of polygons & \% of polygons & \# of polygons & \% of polygons \\ \hline
0\% to $\pm$20\% & 1135 & 67\% & 7 & 0\% \\
+20 to +100\% & 138 & 8\% & 12 & 1\% \\
-20 to -100\% & 268 & 16\% & 27 & 2\% \\
\textgreater{}100\% & 22 & 1\% & 50 & 3\% \\
Non-Applicable & 140 & 8\% & 1607 & 94\%
\end{tabular}%
\caption{Relative error of the population estimates using WorldPOP and LandScan gridded population data.}
\label{Grids}
\end{table}
\section{Conclusions and Future Work}
In this paper, we assess the quality of two gridded population data (WorldPOP and LandScan) for estimating the population living in slums. Whilst LandScan did not produce satisfactory results, WorldPOP provided an estimate that was just 5.9\% less than the ground-truth data, which suggests that gridded population data with a resolution of at least 100m can be a useful tool for estimating the population living in slums. \par
In the future, we would like to extend this work to countries where the ground-truth was collected in the year 2015, since more gridded population dataset have estimates for that year. Also, we would like to contemplate regions whose census is shared at a lower granularity than Brazil, as these countries may have less accurate gridded population estimates. Ideally, we would like to compare the results of top-down approaches, such as the population layers described in this paper, with bottom-up ones, such as the Geo-Referenced Infrastructure and Demographic Data for Development (GRID3), currently only available for four countries, and the World Settlement Footprint 2015, which is at the moment under restricted access. Finally, to achieve goal 11.1 of the Sustainable Development Goals, the analysis pipeline developed in this study needs to be extended to include a cost-effective method to detect slum areas and the conditions in these settlements improved. 
\section*{Broader Impact}
This research raises many ethical issues that should be considered. Firstly, it is important to be aware of the many definitions of the word “slum” \cite{lilfordBecauseSpaceMatters2019} and that, despite it being used in the Sustainable Development Goals and United Nations Habitat reports \cite{habitatTrackingProgressInclusive2018a}, it can be considered a derogatory term \cite{doveyViewpointInformalSettlement2020}. Secondly, machine learning applications, like gridded population data and scripts to extract information from these applications (such as the methodology proposed in this paper) can be useful to inform public policy and monitor progress towards goals, but they do not substitute field-surveys and interaction with the affected communities. In fact, our research will be of best use if adopted in conjunction with official census data collection to address quality issues that might arise in the data collection process \cite{mahabirCriticalReviewHigh2018a} or the reluctance of some governments to include slums in official statistics \cite{habitatTrackingProgressInclusive2018a}. Finally, for the research that quantifies the population living in deprived areas to promote positive social impact, its output needs to be used with the interest of the poor in mind.\par
\section*{Acknowledgments}
Science Foundation Ireland supported this research through the SFI Centre for Research Training in Machine Learning (18/CRT/6183).

\small
\bibliography{ML4D20refAgatha}

\end{document}